**Title:**

Is the free-energy principle a formal theory of semantics? From variational density dynamics to neural and phenotypic representations

**Authors:**
Maxwell J. D. Ramstead [1,2,3]
Karl Friston [3]
Inês Hipólito [4,5]

**Affiliations:**
1. Division of Social and Transcultural Psychiatry, Department of Psychiatry, McGill University, Canada.
2. Culture, Mind, and Brain Program, McGill University, Canada.
3. Wellcome Centre for Human Neuroimaging, University College London, United Kingdom.
4. Faculty of Arts, Social Sciences, and Humanities, University of Wollongong, Australia.
5. Institute of Psychiatry, Psychology and Neuroscience (IoPPN), King's College London, United Kingdom.

**Abstract:**
The aim of this paper is twofold: (1) to assess whether the construct of neural representations plays an explanatory role under the variational free-energy principle and its corollary process theory, active inference; and (2) if so, to assess which philosophical stance—in relation to the ontological and epistemological status of representations—is most appropriate. We focus on non-realist (deflationary and fictionalist-instrumentalist) approaches. We consider a deflationary account of mental representation, according to which the explanatorily relevant contents of neural representations are mathematical, rather than cognitive, and a fictionalist or instrumentalist account, according to which representations are scientifically useful fictions that serve explanatory (and other) aims. After reviewing the free-energy principle and active inference, we argue that the model of adaptive phenotypes under the free-energy principle can be used to furnish a formal semantics, enabling us to assign semantic content to specific phenotypic states (the internal states of a Markovian system that exists far from equilibrium). We propose a modified fictionalist account – an *organism-centered fictionalism or instrumentalism*. We argue that, under the free-energy principle, pursuing even a deflationary account of the content of neural representations licenses the appeal to the kind of semantic content involved in the 'aboutness' or intentionality of cognitive systems; our position is thus coherent with, but rests on distinct assumptions from, the realist position. We argue that the free-energy principle thereby explains the aboutness or intentionality in living systems and hence their capacity to parse their sensory stream using an ontology or set of semantic factors.

**Acknowledgments:**
We are grateful to Alex Kiefer, Ryan Smith, Conor Heins, Wanja Wiese, Krys Dolega, Mel Andrews, Casper Hesp, and Alec Tschantz for helpful comments and discussions that

 

contributed to our work on this paper. Researchers on this article were supported by the Social Sciences and Humanities Research Council of Canada (MJDR), by a Wellcome Trust Principal Research Fellowship (Ref: 088130/Z/09/Z) (KJF), and by the University of Wollongong (IH).

 

## 1. Introduction: Neural representations and their (dis)contents

Representations figure prominently in several human affairs. Human beings routinely use representational artifacts like maps to navigate their environments. Maps represent the terrain to be traversed, for an agent capable of reading it and of leveraging the information that it contains to guide their behavior. It is quite uncontroversial to claim that human beings consciously and deliberately engage in intellectual tasks, such as theorizing about causes and effects—which entails the ability to mentally think about situations and states of affairs. Most of us can see in our mind's eye situations past, possible, and fictive, via the imagination and mental imagery, traditionally cast in terms of representational abilities.

In the cognitive sciences, neurosciences, and the philosophy of mind, the concept of representation has been used to try and explain naturalistically how the fundamental property of 'aboutness' or 'intentionality' emerges in living systems (Brentano, 1973). Indeed, living creatures must interact with the world in which they are embedded, and must distinguish environmental features and other organisms that are relevant for their survival, from those that are not. Living creatures act as if they had beliefs about the world, about its structure and its denizens, which guide their decision-making processes, especially with respect to the generation of adaptive action. This property of aboutness is thus a foundational one for any system that must make *probabilistic inferences* to support their *decision-making* in an uncertain world, which are central to the special issue to which this paper contributes.

In this context, to provide a naturalistic explanation is to explain some phenomenon by appealing to physical processes (Haugeland, 1990). The strategy deployed by cognitive science has been to naturalize intentionality by postulating the existence of physical structures internal to the agent that carry, encode, or otherwise bear semantic content; classical accounts include Fodor (1975); Millikan (1984, 1989).

In the sciences that study the mind and in philosophy, representations are typically defined as some internal vehicles – *neural representations* – that carry semantic content (Ramsey, 2007). Representations are thus physical structures that are internal to an organism; typically, states and processes unfolding in their brains, which carry or encode representational content. The epistemic role played by neural representations is to explain how creatures are able to engage with relevant features of their environment and plan situationally appropriate, adaptive behavior (Boone and Piccinini, 2016; Ramsey, 2007). The semantic content of a representation is what the representation is about, that towards which it stands in an intentional relation – that "in virtue of what they represent what they do, or get to be 'about' what they are about" (Kiefer and Hohwy, 2018, p. 2390). The problem of specifying the nature and origins of semantic content is known as the hard problem of content (Hutto and Myin, 2017; Hutto and Satne, 2015).

There are several well-accepted constraints for the appropriateness of representational explanations: such an account should (1) cohere broadly with the actual practices that are used in computational cognitive science research; (2) allow for misrepresentation, i.e., the representation must be able to "get it wrong"; (3) provide the principled method for attributing of determinate contents to specific states or structures (typically internal to the system), and finally (4) be naturalistic, meaning that the account of semantic content does not itself appeal to semantic terms when defining how the representational capacity is realized by the physical system, on pain of circularity in reasoning (Egan, 2019; Sprevak, 2019).

                                                       

A fundamental and thorny question is whether there is some construct of representation that not only applies in all cases (i.e., to the construct of neural representations and to our more familiar, deliberate, everyday representational activities), but that also really *explains* the intentional relation between living creatures and their environment. An equally thorny issue involves the ontological and epistemological status of neural representations: Do such things really exist? Do they have explanatory value?

The aim of this paper is twofold. First, we aim to determine whether or not the construct of neural representations plays an explanatory role in an increasingly popular framework for the study of action and cognition in living systems, namely, the variational free-energy principle and its corollary process theory, active inference. Second, if the postulation of neural representations is warranted under the free-energy principle, we aim to assess which of the available philosophical positions about the ontological and epistemological status of representations is most appropriate for the construct under this framework. Since the issue to be determined cannot merely be decided by appeal to formal frameworks, we first discuss the issue of representationalism.

In the remainder of this first section, we review the issues surrounding representationalism. In the second section, we present and motivate the view of the brain under the free-energy principle, as a self-organizing nonequilibrium steady-state enshrouded by a statistical boundary (called a Markov blanket). In the following section, we consider non-realist accounts of neural representation: a deflationary account, according to which the contents of neural representations are mathematical, and a fictionalist account, according to which representations are scientifically useful fictions. In the fourth section, we propose to combine aspects of both these accounts, yielding a nuanced realist account that defines semantic contents of representations formally – what one might call a *deflationary*, *organism-centered fictionalist* interpretation of neural representations. We argue that even pursuing a minimalist, deflationary account of the content of neural representations under the free-energy principles licenses an appeal to a robust kind of semantic content, the kind at stake in the 'aboutness' or intentionality of cognitive systems. The ensuing position coheres broadly with, but rests on distinct assumptions from, the realist one.

## 1.1. The faces of representationalism: Realism and non-realism

In the philosophy of mind, there are, roughly speaking, two main ways to think about the ontological status of the neural representation construct, which have implications for the available epistemological positions. One is *realism* about neural representations. This view combines two positions: ontologically, that neural representations really exist (typically, that they are physically instantiated in the brain); epistemologically, that they are scientifically useful postulates as well (Kiefer, 2017; Ramsey, 2007). *Non-realist* positions also are available, which do not share all of these assumptions. Non-realists are either agnostic about the reality of neural representations or explicitly reject the assumption. *Anti-realism* says that neural representations do not exist. Several varieties of non-realism are available, which have different epistemological implications. *Eliminationism* is the anti-realist view that the construct of neural representation should be eliminated from scientific practice (Hutto and Myin, 2013; Thompson, 2010). *Instrumentalism* or *fictionalism* is the non-realist view that argues that neural representations are useful fictions: they are a scientifically useful way of describing the world (Dennett, 1989; Egan, 2019; Egan, 2018; McGregor, 2017; Sprevak, 2013, 2019).

To get clear on which of these positions is most appropriate, it is useful to review the different versions of the neural representation construct. The classical view of neural representations casts them as symbolic structures that are realized by brain states and that are manipulated by rule-governed processes. This follows from the computational theory of mind (Fodor, 1975), according to which cognition is the rule-governed manipulation of symbol-like, internal cognitive structures (i.e., neural representations). In these classical accounts, the content of a representation is determined either by appealing to an innate stock of concepts and mechanisms that ensure the accuracy and objectivity of what is represented (Fodor, 1975; Fodor and Pylyshyn, 1988); by accounting for contents through the actions of a biological proper function (Millikan, 2017; Shea, 2018; Shea et al., 2018); or by referring to the phenomenal content of our first-person experience of things in the world (Horgan and Graham, 2012; Macpherson, 2012). All such accounts have in common a construct of neural representation as an internal symbol (or type) that gets instantiated (or tokened) in the appropriate circumstances; what varies is how the appropriateness condition gets implemented. There are also non-representational versions of the computational theory of mind, which will not concern us here; see Miłkowski (2013); Piccinini (2015).

Motivated by parallel distributed processing, connectionist models of neural representations disagree with proponents of the classical approach over the nature of the representational vehicle; but agree that the brain harnesses internal cognitive structures that act as the vehicles for content (Chalmers, 1993; McClelland et al., 1986). Rather than discrete symbolic structures, the connectionist argues that neural representations are distributed representations; that is, that they accomplish their function of representing states of affairs in terms of joint configurations of their activity.

Today, the most popular (and in our view, the most compelling) representationalist accounts are of the connectionist type. They cast neural representations as *structural representations*. On this account, neural representations are able to represent their target domain (i.e., to encode semantic content about their target domain) because their neural vehicles encode *exploitable structural similarities* shared with the target domain (Kiefer and Hohwy, 2018, 2019; O'Brien and Opie, 2004; Williams and Colling, 2017). On this account, representations function much like maps: they recapitulate the high-order structural features of that domain, for example, its statistical properties; also see (Goh et al., 2018). More specifically, structural representations encode information in a format that the organism can exploit to guide its behavior, that can afford the detection of errors (i.e., that affords misrepresentation), and that can be used for 'offline' navigation (Gładziejewski, 2016; Gładziejewski and Miłkowski, 2017; Hohwy, 2014; Kiefer and Hohwy, 2018, 2019). Structural representations operate iconically, via a process "in which the structure of internal representations in the brain come to replicate the structure of the generative process by which sensory input impinges upon it" (Williams and Colling, 2017, p. 1962).

## 1.2. Towards anti-realism: deficiencies of the realist view

Most research in the cognitive sciences and neurosciences tacitly operate on *realist* assumptions about neural representation and design experiments that aim at explaining how our experience of, and action in, the world is mediated by structured bodies of knowledge that are encoded in the networks of the brain. The existence and explanatory value of neural representations is a basic premise of almost any psychology textbook. For instance, one can read in *The MIT Encyclopedia of the Cognitive Sciences*: "Psychology is the science that investigates the *representation* and *processing of information*" (Wilson and Keil, 2001, p. xl, emphasis added).

Although *realism* about mental representation is the default mode of operation in most of the cognitive sciences and neurosciences, it is not the consensus position. The motivation for adopting an *anti*-realist, *anti*-representationalist approach comes from the observation that, despite enormous efforts and scientific investment, representations have yet to be naturalized (Chemero, 2009; Hutto and Myin, 2017; Kripke, 1982; Loewer, 1997; Putnam, 1981; Quine, 1969; Ramsey, 2007). This is notably because extant attempts to articulate a theory of neural representation have so far failed to provide a naturalistic theory of semantic content that does not presuppose the very intentional relation and representational content that it seeks to explain, securing point (4) discussed above. Thus, scholars such as Ramsey (2016) call for caution about the use of the construct. They argue that often, the appeal to semantic content is a philosophical gloss that does not add any explanatory value:

> "The roles provided by commonsense psychology are those that distinguish different types of mental representations. What we need and what is not provided by commonsense psychology is, more generally, *the sort of physical condition that makes something a representational state*, period. In functional terms, we would like to know what different types of representations perhaps have in common, qua representation. *Neither commonsense psychology nor computationalism tells us much about the sort causal/physical conditions* that bestow upon brain states the functional role of representing (at least not directly)." (Ramsey, 2016, p. 6, emphasis added)

In the literature, it is quite common to see the selective responsiveness of neural tissue to a given stimulus described as the representation or encoding of stimuli. This conceptualization is adopted in the study of perception, particularly to highlight cellular specialization in detecting certain features of the perceptual object (Bonnen et al., 2020; Díaz-Gutiérrez et al., 2020; Hubel and Wiesel, 2004; O'Sullivan et al., 2019; Shadlen and Newsome, 1994); in the study of memory (Kiyonaga et al., 2017; Manohar et al., 2019) and motor activity (Anderson et al., 1990; Gregory, 2018; Stengel, 1994) as well. However, as Ramsey (2007) notes, response-selectivity by itself cannot make a physical state a representation. Many physical states have response-selectivity but are not representations. For instance, the states of one's skin vary with the weather (e.g., it gets darker the more it is exposed to the sun), but we would not (intuitively) count one's skin as a representation of the sun or weather.

We find approaches in the biological sciences and in the neuroscience less committed to the view of cognition as a representational process taking place within the boundaries of the brain (Calvo and Friston, 2017; Engel et al., 2016; Hipolito and Kirchhoff, 2019; Ramstead et al., 2019b; Williams, 2018). These views include perceptual and motor control theory (Arsiwalla et al., 2019; Rosenbaum, 2009); robotics (Beer, 2000); cybernetics (Pickering, 2010; Porush, 2018; Seth, 2014; Von Bertalanffy, 1950); and, arguably, the free-energy principle and active inference (Engel et al., 2016; Friston, 2020; Ramstead et al., 2019b). These accounts, which often hail from embodied and enactive approaches in cognitive science (Gallagher, 2020; Newen et al., 2018; Noë, 2004; Thompson, 2010; Varela et al., 1991), converge on the idea that the primary aim of cognition is not internally reconstructing proxies for the structure of a hidden world, but rather to adapt to and act in an environment.

In this setting, Sprevak (2013) suggests that there appear to be two options. We can take a hard-headed *realist* assumption that the naturalization of neural representations will eventually succeed. (Even though no account has succeeded so far.) Alternatively, *non-realism* (eliminativism and fictionalism) downplays the value of representation talk in cognitive science. Non-realists observe it is not possible to define the content of a neuronal representation without conflating it already with the cognitive process: We borrow our semantics (and thereby, the content of the representation) from our scientific practice. More precisely, if the supposed representational content cannot be determined without appealing to what we know about the cognitive activity itself, then it is the cognitive activity that has explanatory power. Conversely, if the explanation in terms of the cognitive activity suffices without appealing to an experimenter-imposed semantics, then there is no reason to postulate representational content. This may motivate restraining the use of neural representation use to an "informal gloss" (Egan, 2019; Egan, 2018).

The cost of eliminating the construct of neural representation altogether is that it requires a painful revision of the mainstream representational paradigm in cognitive science. Indeed, we typically appeal to neural representations to explain *goal-directed*, *probabilistic inference*, and *decision-making*. To abandon this posit leaves us with the obligation to abandon some of the most powerful explanatory tools that we can use. Is this legitimate?

## 1.3. Representations under the free-energy principle?

The object of this paper is more precisely the status of neural representations in Bayesian neuroscience, known as the free-energy principle and active inference. In Bayesian neuroscience, the brain is cast as a statistical organ or inference engine that minimizes its uncertainty about the state of the world. In this family of theories, the brain is depicted as doing its predictive work by drawing on probabilistic knowledge about its environment to explain the likely causes of the sensory signals with which it is bombarded (Hinton and Sejnowski, 1983; Hohwy, 2014; Rao and Ballard, 1999; Seth, 2014)—and to act in ways that bring about its preferred or expected sensory states (Clark, 2015; Hohwy, 2020).

Some Bayesian approaches based on predictive coding algorithms for describing canonical microcircuits in the cortex have strong representationalist commitments (Clark, 2015; Gładziejewski, 2016; Gładziejewski and Miłkowski, 2017; Hohwy, 2014, 2016; Kiefer and Hohwy, 2018, 2019; Williams and Colling, 2017). These accounts argue that the Bayesian brain entails the postulation of structural representations. Thus, neural representations are taken to be internal, map-like structures that are instantiated in the networks of the brain and that encode exploitable information.

Several recent papers have discussed whether a realist interpretation of neural representations is warranted under the free-energy principle (Gładziejewski, 2016; Gładziejewski and Miłkowski, 2017; Kiefer and Hohwy, 2018, 2019; Ramstead et al., 2019b). Besides a few notable exceptions (van Es, 2020), few papers have sought to evaluate the variety of non-realist arguments in light of the free-energy principle. In this paper, we will argue that taking seriously two forms of the non-realist position (the deflationary and fictionalist views) leads back to a nuanced form of realism that is apt to provide a naturalistic basis for the study of intentionality.

Why concern ourselves with the ontological and epistemological status of representations? From our point of view, one main reason to do so is that one's position with regards to the status of representations has implications for research in computational neuroscience; and to determine which structures play the role of representation, and how they carry their semantic contents, is crucial for the practice of neuroscience. One salient example is that of motor representations and motor commands in the human brain (Friston, 2011b; Hipolito et al., 2020). Representationalist frameworks in computational neuroscience assume that there exist structures in the brain that represent motor tasks. Optimal control theory is one of the more popular frameworks for modelling that borrows such assumptions. This approach is underwritten by strong assumptions about the nature of the models and signals that the brain processes in motor control. The hypothesis in optimal control is that motor representations are brain structures that encode explicit instructions to perform a task and that are specified in terms of intrinsic coordinates (i.e., in terms of the contraction and stretching of muscle fibres). However, optimal control theoretic constructs have been criticized and empirical evidence is lacking for explicit instruction-like motor commands in the brain. For discussion, see Hipolito et al. (2020). Our framework offers an alternative characterization of representational capacities, which are not premised on instruction-like motor commands. Instead, we characterize the representational capacity as underwritten by an ontology of fictive states, and on a process of active inference that realizes preferences about sensory data. Our deflationary perspective on representational capacities does away with the problematic representational posits of optimal control, while also shedding light on how semantic contents are acquired through histories of active inference.

## 2. The free-energy principle and active inference: From information geometry to the physics of phenotypes

### 2.1. State spaces, nonequilibrium dynamics, and bears (oh my)

Our paper will focus on a prominent Bayesian theory of action and cognition, the variational free-energy principle, and its corollary, active inference (Friston, 2010, 2020). The free-energy principle starts with the observation that biological systems like living creatures have a phenotype. Living organisms maintain their phenotypic integrity and resist the tendency towards thermodynamic equilibrium with their ambient surroundings – that is usually dictated by the fluctuation theorems that generalize the second law of thermodynamics (Kiefer, 2020; Parr et al., 2020). Living creatures do so by upper bounding the entropy (the dispersion or spread) of their constituent states. To get a better handle on this, in this section, we introduce two formal notions: the state space and nonequilibrium steady states. A brief technical treatment of what follows can be found in the Technical appendix and Glossary of terms and expressions.

In physics, equilibrium and nonequilibrium are distinguished by the end-state towards which a dynamics evolves. Equilibrium dynamics resolve when all the energy gradients have been consumed; at which the point system is at thermodynamic equilibrium with its environment. For living creatures, thermodynamic equilibrium is death. Living creatures are open systems that remain far from equilibrium. How can we model this using formal resources?

 

The state or phase space formalism comes from dynamical systems theory and allows us to get a formal grip on the predicament of living systems. A *state or phase space* is an abstract space that allows us to model the time evolution of a system in terms of all the possible states in which it can find itself. To construct a state space, we identify all the relevant quantities that can change in the system (i.e., all the relevant variables) and then plot each variable on a dimension in an abstract space. This space is called a state space. Every dimension of this space corresponds to a variable in the system; such that a point in this space corresponds to a complete instantaneous specification of the system, since we assign a value to every variable of the system; i.e., we assign a position to the system along every dimension. A *trajectory* in this space, in turn, corresponds to the flow of the states of the system over time.

The state space formalism allows us to describe the time evolution of a system implicitly by depicting trajectories through state space. This turns out to be crucial. If we draw a probability density over all the states that a system can find itself, those combinations of states with the highest probability, to which the system returns periodically, are known as a pullback attractor (Crauel and Flandoli, 1994; Friston, 2020). We can associate the states (i.e., the regions of this space) in which a creature finds itself most of the time with its *phenotypic states*.

The probability density that describes the system at its nonequilibrium steady state (i.e., its phenotypic states) is aptly called the *nonequilibrium steady state density*. Such a probabilistic description of the system's dynamics can be interpreted in two mutually consistent and complementary ways. First, the system can be described in terms of the flow of the system's states—that are subject to random fluctuations—in which case, we can formulate the flow in terms of a path integral formulation, as a path of least action. Equivalently, we can describe the nonequilibrium steady-state in terms of the probability of finding the system in some state when sampling at any random time. These two descriptions are linked mathematically by the fact that at nonequilibrium steady-state, the flow is the solution to something called the Fokker Planck equation that describes the density dynamics. This dual interpretation will play a crucial role later.

## 2.2. Markov blankets and the dynamics of living systems

The free-energy principle builds on the dynamic systems theoretic approach, which concerns the time evolution of systems, but now augmented with considerations about the statistical properties and the measurability of those systems. The free-energy principle allows us to describe the flow of a system's states in one of two mathematically equivalent ways, statistical and dynamical – an equivalence that is warranted by virtue of the conditional independence entailed by the presence of a statistical boundary (called a Markov blanket) and the existence of the system in a regime of steady (phenotypic) states far from equilibrium.

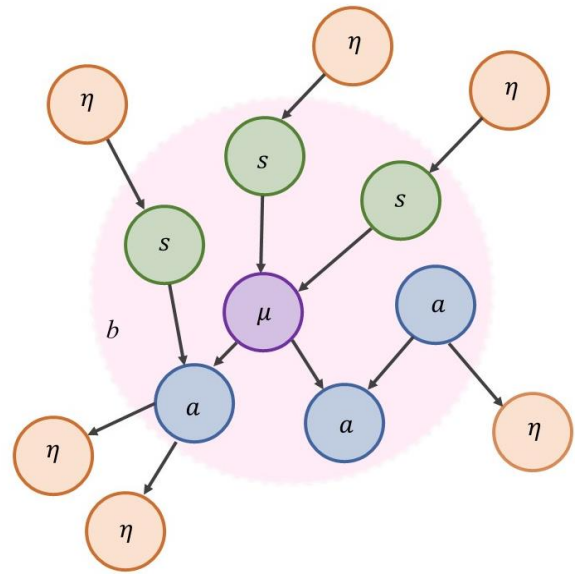

**Figure 1.** *The structure of a Markov blanket.* A Markov blanket highlights open systems exchanging matter, energy, or information with their surroundings. The figure depicts the Markovian partition of the system or set of states into internal states (denoted $\mu$), blanket states, which are themselves divided into active states ($a$) and sensory states ($s$) states, and external states ($\eta$). Internal states are conditionally independent from external states, given blanket states. Variables are conditionally independent of each other by virtue of the Markov blanket. If there is no route between two variables, and they share parents, they are conditionally independent. Arrows go from 'parents' to 'children'. From Hipólito et al. (2020).

For a system to exist (separately from the rest of the universe), it must be endowed with a degree of independence from its embedding environment. A Markov blanket is a set of variables or states that we use to stipulatively individuate a system in terms of what is part of it (internal states, denoted $\mu$), and what is not (external states, denoted $\eta$). The Markov blanket itself is defined as those states that mediate interactions between the system and its embedding environment (active and sensory states, denoted $a$ and $s$). The Markov blanket is defined by the absence of certain connections: internal states do not cause sensory states, and external states do not cause active states. See Figure 1 and Figure 2.

The presence of a Markov blanket induces a conditional independence between internal and external variables. The key word here is 'conditional': internal and external are not really independent of one another—they just appear to be so if we discount their dependencies via the blanket states.

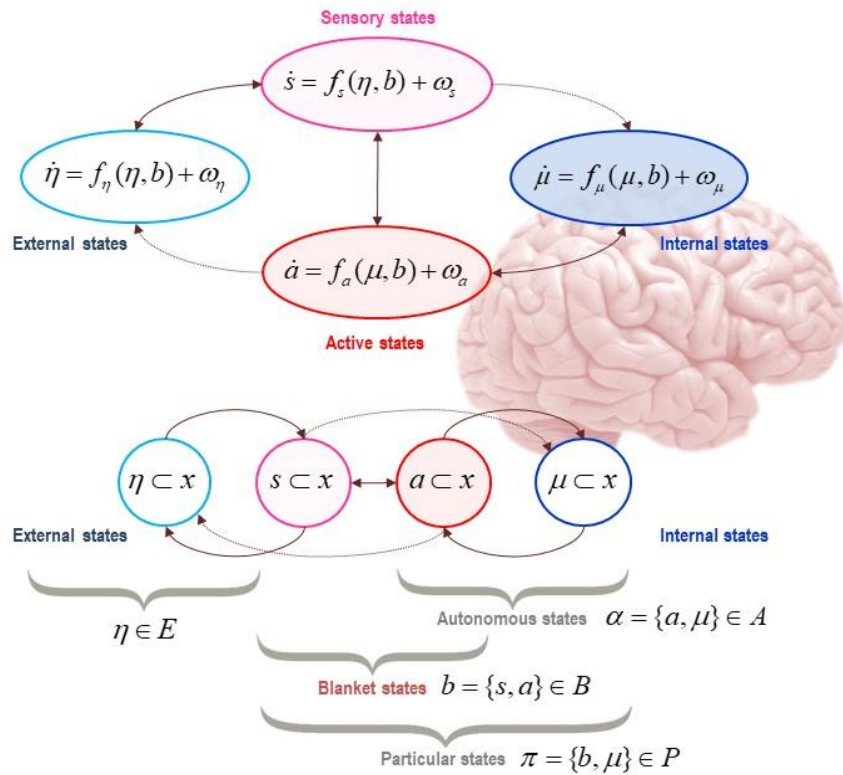

**Figure 2.** *Markov blankets of life.* This figure depicts a Markov blanket around a system of interest; here, the brain. The figure associates the Markovian partition to internal ($\mu$), blanket ($b$), and external states ($\eta$); where blanket states can be active ($a$) or sensory ($s$) states, depending on their statistical relations to internal and external states. Here, the flow of each kind of state (denoted with a dot) is expressed as a function of other states in the partition (plus some random noise, denoted $\omega$), as a function of the independences that are harnessed by the Markov blanket. Autonomous states ($a$) are those that follow a free-energy gradient. Particular states (here, denoted $\pi$) are those identified with the system itself (internal and blanket states).

## 2.3. Information geometries and the physics of sentient systems

The core intuition behind the free-energy principle is that if a system is endowed with a phenotype (i.e., a nonequilibrium steady state density) and has a Markov blanket, then there are two ways of describing the flow of a system's states that turn out to be equivalent: one rooted in the state space description of the system that is formulated in terms of the flow or dynamics of the internal states; the other rooted in a statistical interpretation of the same flow. The presence of a Markov blanket in such a system ensures that both descriptions are coincidentally true or conjugate to each other (Friston, 2020; Friston et al., 2020).

This description of the system—in terms of movement in internal phase space—is the system's 'intrinsic' information geometry, which is closely related to measure theory and statistical thermodynamics. Measure theory is a field of mathematics about how we can systematically assign a number to subsets of a given set, where a measure or metric consists precisely in such an assignment; for example, a probability measure assigns a probability value systematically to elements of a subset. We can think of measures as capturing something about size, or distances in an abstract space. Generally, an arbitrary set of points that compose a state space does not have a measure, and so no associated notion of distance can be defined for that space. However, one can equip a space with a metric, usually in the form of a matrix that describes how 'far' one has moved as a function of small changes in position.

In Euclidean geometry, this metric is just an identity matrix, i.e., if I move 100 meters along some particular direction, then I will have moved a total of 100 meters. This is not true, e.g., of movement on a sphere: the planet Earth, for instance, has a circumference of approximately 40,000 km; if I move (approximately) 40,000 km in the same direction on Earth, I will not have moved at all in total relative to the Earth's surface (since I will be back where I started).

This notion of metric plays a special role when dealing with sufficient statistics and statistical manifolds. *Sufficient statistics* are the minimal set of numbers that are needed to reconstruct or parameterize a probability distribution (which contains an infinite number of points). For normal or Gaussian distributions, these numbers are the mean and variance. A *statistical manifold* is a space in which the coordinates are the sufficient statistics of a family of probabilities densities. For instance, the sufficient statistics of a Gaussian distribution are its mean and variance, giving a two-dimensional statistical manifold or state space. Given any position on this manifold, it is possible to reconstruct the probability density (which assigns a probability to an infinite number of points). Trajectories on a statistical manifold correspond to changes in the shape (i.e., the value of the parameters) of the associated probability density.

*Information geometry*, in turn, is the field of mathematics dealing with measures or metrics on a *statistical manifold* (information manifolds, typically endowed with a Fisher or Riemannian information metric). In other words, information geometry allows us to define a metric for probability distributions (or probability densities for continuous variables); that is, we can talk about distances between probability densities. Furthermore, if we associate probability densities with probabilistic beliefs, we now have a naturalized way of talking about the distance between beliefs.

With all this in place, we can appreciate what the free-energy principle brings to the table. Starting from our state space description of the system, we can define a metric that allows us to speak about distances between probabilistic configurations of a system's internal states. Such a geometry is 'intrinsic' because it describes the structure of a system's possible configurations in a manner that only refers to internal states themselves (rather than the external states of the environment). The free-energy principle says that so long as the system at hand is equipped with a nonequilibrium steady state density and a Markov blanket is in play, then an additional—and mathematically conjugate—information geometry can be defined (Friston, 2020; Friston et al., 2020). These two information geometries can take the form of the following (Fisher information) metrics, where the sufficient statistics correspond to expected thermodynamic states and internal states, respectively:

$$g(\lambda) = \nabla_{\lambda'\lambda'} D[\, p_{\lambda'}(\mu) \,\|\, p_{\lambda}(\mu)\,]\big|_{\lambda'=\lambda} \quad \text{intrinsic}$$

$$g(\mu) = \nabla_{\mu'\mu'} D[\, q_{\mu'}(\eta) \,\|\, q_{\mu}(\eta)\,]\big|_{\mu'=\mu} \quad \text{extrinsic}$$

This licenses a spectacular observation: namely, that internal states can be interpreted in terms of their extrinsic geometry, i.e., as parameterising a *probability density* over external states. This simple fact is a natural consequence of the conditional independences that define the Markov blanket. Put simply, for every blanket state (i.e., joint sensory and active state) there is a conditional probability density over internal and external states. Crucially, these are conditionally independent, by definition, given the blanket state in question. This means that for every expected internal state, given the blanket state, there must be a conditional probability density over external states. In turn, this means that the expected internal state is a statistical manifold—equipped with an extrinsic information geometry. This extrinsic information geometry describes the distance among probabilistic beliefs *about external states*, which are parameterized by the expected internal states. In other words, expected internal states constitute the sufficient statistics of beliefs about external states.

## 2.4. Phenotypes: A tale of two densities

Essentially, the free-energy principle is a story about two probability densities (Ramstead et al., 2019b). The first is the nonequilibrium steady state density itself, which harnesses the statistical structure of the phenotype. The second, *variational density*, is parameterized or embodied by the internal states of the system. We have seen that internal states constitute the sufficient statistics of probabilistic beliefs about external states. Another way of looking at this is to say that internal states encode a probability distribution—the variational density—over the states of the external world that are generating sensory (and active) states. Technically, that a system will evolve towards—and maintain—its nonequilibrium steady state means that it minimizes the discrepancy between the variational density that it embodies and the probability density over external states, given blanket states (Friston, 2020; Friston et al., 2020). This discrepancy is the variational free-energy, and the steady state flow that underwrites nonequilibrium steady-state becomes a gradient flow on variational free-energy. In other words, it will look as if internal states are trying to optimize their posterior beliefs about external states. When cast in terms of movement in the extrinsic geometry, one can interpret existential behavior in terms of Bayesian belief updating. This all follows because the expected internal states parameterize a conditional or Bayesian belief about external states.

Anthropomorphically, the system does not 'know' in what state it is in, but it will look as if it is inferring the state of its external milieu 'out there', based on prior beliefs and its current sensory states. The state of the external world is thus never directly perceived, it is instead something that the organism infers and brings about actively through interactions with the world. In this sense, the implicit inference is enactive, in the pragmatist sense of being for action (Anderson, 2017; Engel et al., 2016; Ramstead et al., 2019b). When the organism interacts with the world, it perturbs external states and consequent sensory states. These sensory impressions couple back to internal states that attune to the world around them. Put otherwise, the organism engages in a form of Bayesian inference (i.e., active inference), with respect to the state of its ecological environment, based on the situated interaction.

Why inference? The free-energy principle says that living things exist in virtue of gradient flows on an information theoretic quantity called *surprisal*. This is the solution to the Fokker Planck equation that furnishes nonequilibrium steady-state. Crucially, the dynamics of internal (and active) states at nonequilibrium steady-state can be cast equivalently as flowing down surprisal or free-energy gradients. Free-energy scores the atypicality of sensory (and active) states, given a (generative) model of how those data was generated (Bogacz, 2017; Buckley et al., 2017; Friston, 2020).

The equivalence between surprisal and free-energy rests upon the fact that the (expected) internal states parameterize beliefs about external states. When this variational density corresponds to the density over external states, conditioned on blanket states, surprisal and free-energy are the same. Crucially, free-energy is a functional (i.e., function of the function) of the variational density and an implicit *generative model*. The generative model is just the nonequilibrium steady-state density over external and blanket states. And can be regarded as a description of how external states generate blanket states.

In what follows, we will treat the generative model as an implicit attribute of any nonequilibrium steady-state that possesses a Markov blanket. The generative model is implicit because the only thing needed to describe self-organization and belief updating are the free-energy gradients. This means that the free-energy and its generative model are not evaluated or realised explicitly. This is sometimes referred to as *entailing* a generative model. (Friston, 2012; Ramstead et al., 2019b). In short, the free-energy is a functional of two densities, the generative model and the variational density encoded by internal states.

For people familiar with information theory, surprisal is also known as self-information, where the long-term average of self-information is entropy. This means that nonequilibrium steady-state flows counter the entropy production due to random fluctuations. In turn, this means that the kind of inference implicit in the flow of autonomous states (namely, internal and active states) underwrites the existential imperative to maintain a steady-state far from equilibrium.

What the free-energy principle says then, is that so long as a Markov blanket is in play at nonequilibrium steady state density, gradient flows on surprisal (a function of states) are equivalent to gradient flows on free-energy (a function of sufficient statistics), where the sufficient statistics parameterize probability distributions over—or beliefs about—external states. This echoes the reasoning above about the conjugate information geometries, in terms of dynamics and in terms of statistics. Gradient flow here just means that the autonomous states (i.e., internal and active states; see Figure 2) flow down variational free-energy gradients. And this, in turn, is just another way of talking about *action and perception*. This scheme, naturalizing action and perception as the gradient flow of active and internal states (respectively) on variational free-energy, is known as *active inference*, the corollary process theory of the free-energy principle.

The main take-home message is that the free-energy principle casts the phenotype in two complementary ways: as a flow of states at nonequilibrium steady state (described via an intrinsic information geometry) and a flow that entails belief updating (described via an extrinsic information geometry). In virtue of the active states, the apparent role of internal states is not merely to infer the causes of sensory data, but to generate appropriate patterns of interaction. This means that internal states could parameterize beliefs about the consequences of action, and facilitate the consequences of action for beliefs (Friston, 2020). Generative models are thus not designed to merely do the interpretative work of determining the true state of the world, they cover the consequences of acting on worldly states.

## 2.5. Living models: A mechanistic view on goal-directed, probabilistic inference and decision-making under the free-energy principle

With this setup in place, we are in a position to appreciate how generative models allow organisms to engage in goal-directed, probabilistic inference, and decision-making under the free-energy principle. The free-energy principle is often cast as the claim that living systems *just are* generative models of their environment (Allen and Friston, 2016; Friston, 2011a; Ramstead et al., 2018; Ramstead et al., 2019a; Ramstead et al., 2019b; van Es, 2020). We can now make sense of this seemingly enigmatic claim. The free-energy principle says that organisms leverage the statistical structure of their acting bodies to remain in their phenotypic states, where that typical structure is interpreted probabilistically, as a joint distribution over all systemic states. Decision-making about what to do next is then based on a probabilistic inference about "what I must be doing, on the assumption that I am a free-energy minimizing creature."

Hitherto, we have considered inference as an emergent property of self-organization to nonequilibrium steady-state. The ensuing Bayesian mechanics is licensed by an equivalence between surprisal and variational free-energy. In what follows, we can make a further move and describe creatures or particles in terms of the generative model that defines free-energy. Once we have the free-energy, we know the gradients. Once we have the gradients, we know the gradient flow. Once we have the gradient flow, we can naturalize any embodied exchange with the environment.

A typical generative model is depicted in Figure 3. Here, hidden states ($\eta$) correspond to the external states that are hidden from the internal states behind the Markov blanket. The crucial thing to appreciate is that beliefs about hidden states correspond to a *hypothesis*—that the organism embodies— about the causes of its sensations. These hidden states arguably have all the properties that would make them the content of structural representations (Constant et al., 2019; Gładziejewski, 2016; Gładziejewski and Miłkowski, 2017; Kiefer and Hohwy, 2018, 2019; Ramstead et al., 2019b; Williams and Colling, 2017). The hidden states of the generative model are parameterized by the internal states of the system (e.g., the brain) and encode exploitable information about external states that guide adaptive behavior. We will return to this point later.

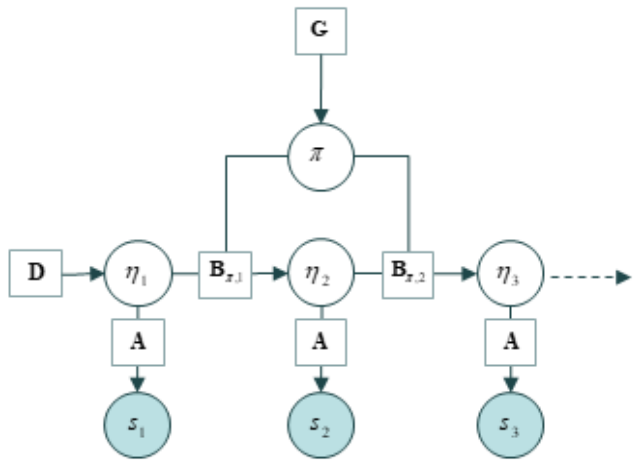

**Figure 3.** *A generative model.* Here, sensory states are observed outcomes, denoted *s*. The generative model represents external states as hidden states, denoted $\eta$. Depicted in this schema are the likelihood mapping from hidden states (denoted **A**), prior beliefs about the probability of state transitions (**B**), and the prior beliefs about initial (hidden) states (**D**). The **G** term is an expected free-energy that drives policy selection ($\pi$) in elaborated generative models that entail the consequences of action (not shown here). The form of this generative model assumes discrete states and steps in time shown from the left to the right. This kind of generative model is known as a hidden Markov model or partially observed Markov decision process.

To summarize, the presence of a Markov blanket at nonequilibrium steady state allows us to associate a living particle or creature with its internal and blanket states. The flow of the internal states acquires a dual aspect, described by conjugate intrinsic and extrinsic information geometries. These geometries inherit naturally from the Markovian structure of the partition. The intrinsic information geometry describes the thermodynamic behavior of internal states (e.g., neuronal dynamics). However, the internal states also are equipped with an extrinsic geometry, which pertains to the probability distributions over, or beliefs about, external states that are parameterized by internal states. A gradient flow of active and internal states on free-energy (action and perception), effectively implements active inference; namely, inferring external states and planning what to do next. This completes our technical review of the free-energy principle.

## 3. Deflationary and fictionalist accounts of neural representation

Having reviewed the technical core of the free-energy principle, we turn to the question of which interpretive frame—from cognitive science—is best positioned to make sense of its representational commitments (or lack thereof). Since realist accounts have been reviewed extensively elsewhere, we focus on two novel non-realist accounts.

### 3.1. A deflationary approach to neural representation

In this section, we examine an interesting position that accommodates aspects of realism and non-realism; namely, the function-theoretic, deflationary account of representation (Egan, 2019; Egan, 2018); for a related but distinct account focused on the contents of the brain, but which does not appeal to information geometry under the free-energy principle, see Wiese (2017). We argue that a suitably amended version of this position yields the best interpretation of representations under the free-energy principle.

A deflationary account of representation (Egan, 2019; Egan, 2018) blazes a path between realism and non-realism: it is realistic about the existence of neural representations as information processing mechanisms that can be characterized using computational methods, but anti-realist about the cognitive contents of these representations. The deflationary account is the view that semantic contents ascribed by scientists merely have an facilitatory role in the explanation of a cognitive capacity; and that whatever aspect of content is explanatorily useful can be specified mathematically. The account builds upon two premises: (1) that representations are not individuated or picked out by their contents, but instead by the mathematical function that it helps to realize; (2) that this content is not essentially determined by a naturalistic relation between states and the structure of the target (i.e., some target domain in the world), but instead by what the mathematical contents are.

Thus, a deflationary account of representations argues (1) that a computational theory of a cognitive capacity must provide a functional theoretic characterization of that capacity, where (2) for the sake of scientific practice, this can be accompanied by an 'intentional gloss' or semantic interpretation, "content is the 'connective tissue' linking the sub-personal mathematical capacities posited in the theory and the manifest personal-level capacity that is the theory's explanatory target" (Egan, 2019, p. 253)

The deflationary account argues that the explanatory work accomplished by representational theories of cognition consists in providing mathematical (functional-theoretic) analyses of a given capacity; and that this computational theory is, more often than not, accompanied by a cognitive or intentional interpretation, which plays a heuristic role, not an explanatory one (Egan, 2019; Egan, 2018). On this account, the contents of a cognitive capacity can be explained naturalistically by appealing to the mathematical functions that are realized by a system. This mathematical content is essential to the computational characterization of the physical process, "if the mechanism computed a different mathematical function, and hence was assigned different mathematical contents, it would be a different computational mechanism" (Egan, 2019, p. 252). This allows for a computational description of the system that is not yet related to a cognitive activity in a certain environment.

Determining content is, on this account, relatively easy, as the mathematical functions deployed in computational models can be understood independently of their use by the system being studied, i.e., independently of the process that is modeled. This happily responds to the most common lacuna of naturalistic representational theories, which is to presuppose the very content they seek to explain by appealing to scientific practice. The characterization of the mathematical content of a neural representation is harnessed in the *computational theory proper*, which is composed of five elements (Egan, 2019; see Figure 3):

1.      A *mathematical function* that is realized by the cognitive system;
2.      *Specific algorithms* that the system uses to compute the function;
3.      *Representational structures* that are maintained and updated by the mechanism;
4.      *Computational processes* that are defined over representational structures.
5.      *Ecological component*: physical facts about the typical operating conditions in which the computational mechanism typically operates.

To this, it is often added a heuristic cognitive content (Egan, 2019). This cognitive content corresponds to observer-based ascriptions of semantics to neural vehicles, often based on reliable covariations between the responsiveness of neural tissue and stimuli presented to that tissue in experimental settings. On the deflationary account, these cognitive contents are an "intentional gloss" on the mathematical characterization provided by the computational theory, which does all the heavy lifting, in explaining the capacity or process under investigation. The environment properties that scientists take to the mechanism to be representing, on this account, are not an essential characterization of the device or computational theory. Rather, they are simply ascribed to facilitate the explanation of a relevant capacity, i.e., they are an intentional gloss on the mathematical content.

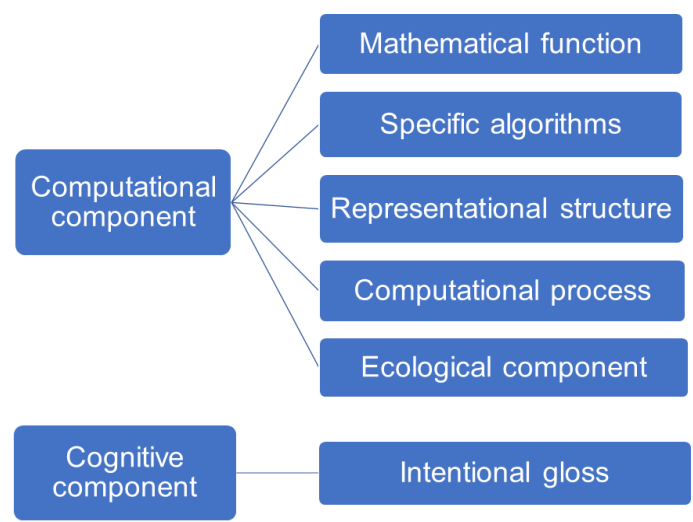

**Figure 4. The deflationary account of the content of a representation.** This figure depicts the main components of the semantic content of neural representations according to the mathematical-deflationary account of content (Egan, 2019). The computational component of the representational content (the computational theory proper) is interpreted in a realist sense. The computational component of the content comprises (1) a mathematical function, (2) specific algorithms that realize this function in the system, (3) physical structures that bear representational contents; (4) computational processes that secure these contents, and (5) normal ecological conditions under which the system can operate. The cognitive content is taken in an anti-realist sense, as a kind of explanatory gloss that only has an explanatory, instrumentalist role, as the interpretation given to the neural representation by the experimenter.

The pressures inherent to the problem of the determination of content brings the proponents of non-realist views to downplay the role traditionally attributed to cognitive content (Egan, 2019; Sprevak, 2019). On these views, and against the dominant view in the cognitive sciences, neurosciences, and philosophy, cognitive content is not the mark of cognition and does not stand in a naturalistic relation between internal states or vehicles of a representation and the structure of the target domain that it models. Specifically, the cognitive components that relate the computational mechanism to actual things in the world are not part of an essential characterization of cognitive activity on the deflationary account; they help define the explanandum but do not figure in the explanation itself. Instead, cognitive or semantic content is merely ascribed by experimenters to facilitate the explanation of a relevant capacity. In a nutshell, what we intuitively take to be the bona fide semantic content of a

                                                19

representation turns out to be a mere gloss that only has a heuristic role in the construction of a scientific explanation.

This runs the risk of trivializing cognitive content—and the intentional relation that it purports to naturalize. The deflationary model does offer an approach to explanation in the sciences of the mind that is perspicuously informative and explanatorily useful to guide research practices in neuroscience. But this comes at a cost. In the deflationary account of content, the role attributed to cognitive content becomes so weakened that it is no longer essential to characterizing the cognitive activity; the only explanatory use of cognitive content is to help scientists systematically make sense of the normal operating conditions in which mathematical descriptions of various mechanisms are deployed. Ultimately, taking the deflationary option seems to undermine the supposed motivation for positing mental representations in the first place, and turns out to dovetail with radical enactivism (Hutto and Myin, 2017; Hutto and Myin, 2013).

### 3.2. Fictionalism and models in scientific practice

Other accounts are available that have a non-realist and perhaps less resolutely *anti-realist* flavor (Salis, 2019; Sprevak, 2013; van Es, 2020). *Fictionalist* or *instrumentalist* accounts in the philosophy of science suggest that scientific models are useful fictions: they are not literally true, but "true enough" or "good enough" to make useful predictions about, and act upon, the world. A fictionalist account of neural representations suggests that they are useful functions used by scientists to explain intentional behavior: they are *models used by scientists*.

In the philosophy of science, model-based approaches (Giere, 1999, 2010; Van Fraassen, 1980) suggest that the work of science consists in the comparison of different sorts of models. The notion of "empirical adequacy" or, heuristically, of the "true enough" occupies a center stage in the debate in epistemology and the philosophy of science (Elgin, 2017; Friend, 2019; Giere, 1999, 2010; Salis, 2019; Van Fraassen, 1980). This notion allows some degree of divergence between what the model postulates and what we find in reality; it entails that models need not be veridical representations of states of affairs. Scientific progress often rests on idealization, and successful models often deliberately contain "felicitous falsehoods" that, while not depicting the world as it "really is," do have value and explanatory power. Examples of this kind of heuristic use of models include the ideal gas model in statistical mechanics and the Hardy–Weinberg model in genetics, both of which occupying central roles in their respective disciplines, but which are not literally true descriptions.

On this account, models play a significant role in the understanding of a subject matter, not despite the fact they do not accurately reflect the world's causal structure, but precisely *because* they are only "true enough" – they allow researchers to focus on the features that are relevant to the hypothesis being tested, by excluding non-relevant features (Giere, 1999, 2010). It is worth noting that modeling is a non-reductive context of inquiry, i.e., a target system that is studied using modeling methods does not have to be reduced to what is modeled (Weiskopf, 2018).

The aim of an explanation is to generate understanding (Elgin, 2017; Frigg and Nguyen, 2017; Grimm et al., 2016). Our appreciation of the explanatory role of models in the practice of science does not depend on a realist interpretation of models (Van Fraassen, 1980). Models are useful, sometimes independently of their capacity to explain a phenomenon. If a model provides explanations that do not accurately represent the causes of their target system, it does not necessarily follow that these explanations are not real explanations (Elgin, 2017). In science, models can be used, for instance, to build new models (Peschard, 2011). There are non-explanatory uses of models, i.e., uses that do not leverage their representational capacities per se (Isaac, 2013). Models can play an explanatory role despite not accurately representing the properties of the target domain (Rice, 2015).

A subtle question is whether the generative models that figure in the free-energy principle and active inference are to be interpreted in a realist or instrumentalist way. That is, are the generative models of the free-energy principle models used by experiments to explain the behavior of cognitive systems, or are such models literally being leveraged by organisms to remain alive and to act adaptively? This ambiguity has been highlighted in a recent paper (van Es, 2020). We turn to this issue next, as we critically amend the deflationist conception of neural representation.

## 4. A variational semantics: From generative models to deflated semantic content

### 4.1. A deflationary account of content under the free-energy principle

In this last section, we combined elements of the deflationary and the instrumentalist accounts of neural representations to propose a kind of *organism-centered fictionalism or instrumentalism*. We expand upon the ecological component of deflated mathematical content, which we argue leads to a naturalistic theory of intentionality: a formal theory of semantic content based on the free-energy principle. The key to formulating a robust mathematical theory of semantic content, one capable of naturalizing intentionality, is to notice that the free-energy principle essentially tells a story about the mutual attunement between a system and its environment.

In our view, the deflationary view of representational contents downplays the role of the fifth, ecological component of the computational theory proper. We argue that formulating the ecological component using the resources of the free-energy principle allows us to salvage the intentionality of semantic content—and thereby recover a robust conception of content tied to the domain to which it is intentionally related (or about)—without appealing to the artificial intentional gloss of cognitive content. The resulting view is of a semantics that emerges naturally from the fact that the system we are considering is equipped with a dual information geometry of states and beliefs.

The formalism that underwrites the free-energy principle licenses a crucial observation: namely, that the mathematical structures and processes in play are defined over a state space and, implicitly, over an associated belief space or statistical manifold (Constant et al., 2019). The mathematical framework of the state space formalism means that the system's dynamics are defined over states of the system; and that as a consequence of the associated extrinsic information geometry, we can always associate a *semantics* to this intrinsic description.

This semantics comes from the 'beliefs' built into the extrinsic information geometry. The term 'belief' is used in the sense of 'belief propagation' and 'Bayesian belief updating', which are just ways of talking about probability distributions or densities. 'Beliefs' in Bayesian terms are posteriors and priors, corresponding to the probability distributions (a world of possible states) that are shaped by physically realized states (i.e., the internal states that parameterize distributions over external states). In general, although we use the term 'beliefs' to describe the probability densities defined over external states, it is generally recognized that these densities are not themselves the same as propositional beliefs. In brief, propositional beliefs have truth conditions; that is, they are the kind of thing that can be true or false (Hutto and Myin, 2013). The probability densities at play here are not of this kind; they represent the manner in which variables covary. This does not imply truth-conditionality, which means that they are non-propositional.

It is often noted that one does not obtain semantic content from mere systematic covariation (Hutto and Myin, 2013; Kirchhoff and Robertson, 2018; Ramsey, 2007). However, this argument can be defeated by noting that, under the free-energy principle, for any living system, there is an implicit semantics at play that is baked into the system's dynamics. Importantly, that is just saying that the system's internal dynamics have a probabilistic aspect (and extrinsic information geometry) that connects it to an embedding system. Via the Markovian partition, we can always associate this trajectory of states on internal (statistical) manifold to a semantics – a formal semantics that falls out of the systemic dynamics and that can be characterized purely mathematically.

Thus, we obtain a bona fide formal semantics from interactions between the system and its context, as well as the histories of environmental interactions between an organism and its niche. What follows from our account is a somewhat more 'realistic' deflationary position, a weakly deflationary position according to which the *content* of a representation is indeed the mathematical function that it realizes, but where this computational theory proper entails an implicit semantics.

## 4.2. From a computational theory proper to a formal semantics

Let us now take stock. We have retained the general description of representational content from the deflationary account. We now use this deflationary model to specify the computational theory proper that leads to a formal semantics via the free-energy principle.

The computational theory proper has five components, which we can map to elements of the free-energy formulation. (1) Under the free-energy principle, the *mathematical function* that is realized by the (gradient flows of the) cognitive system is a free-energy functional that measures the divergence between the posterior and variational densities. (2) The *specific algorithms* that the system uses to compute this function is gradient descent on variational free-energy. (3) Echoing the literature on structural representations, the *representational structures* that are maintained and updated by the mechanism are the internal states of the system. (4) The *computational processes* that are defined over these representational structures, and which update and maintain them, are implemented as active inference. (5) Finally, the *ecological component* is provided by the dual information geometry.

Figure 4 reformulates Egan's (2019) account deflationary account of content in light of the free-energy principle. We amend the deflationary account to highlight that it provides us with a fully naturalistic, mathematical account of the origin of semantic content, in terms of a calculus of beliefs and intentions that is the counterpart of the intrinsic description of the flow of internal states. Note that the external states that figure in the generative model implicitly define the *ecological component* – and this, purely mathematically. This overcomes the problem of *naturalizing intentionality* by appealing purely to well understood mathematical and physical processes and properties.

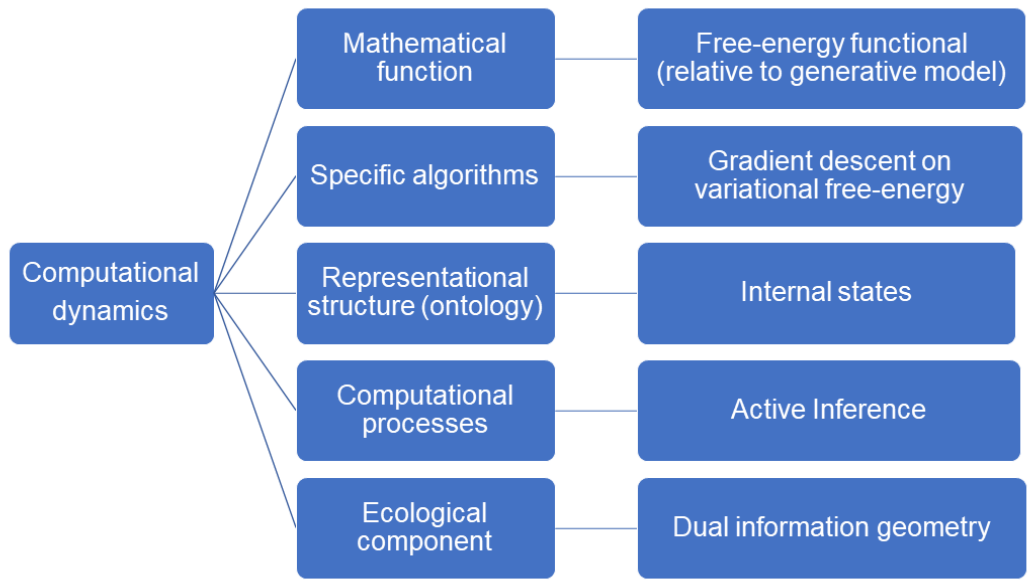

**Figure 5. A formal semantics under the free-energy principle.** Bayesian cognitive science does not have to commit to the classic notion of representation carrying propositional semantic content. The free-energy principle allows us to formulate a formal semantics. "Representations" under the free-energy principle are, in essence, formalized under the physics of flow (e.g., dynamical systems theory) and information geometry, and they are better understood as internal structures enabling the system to parse its sensory stream (i.e., as an ontology). Here, we relate the main components of the deflationary account of content (Egan, 2019) to the free-energy formulation. The mathematical function that underwrites the free-energy principle is a variational free-energy functional. The specific algorithm is gradient descent (i.e. flow) on this free-energy functional, which defines the gradients on which the system 'surfs' until it reaches a nonequilibrium steady state. Representational structures (i.e., the structures that embody or carry out these processes) correspond to the internal states of a system and associated intrinsic information geometry. The computational process itself is active inference, which provides an overarching framework to use the generative model for policy (action) selection. Finally, the ecological component is defined by the implicit semantics that is entailed by the dual (intrinsic and extrinsic) information geometries: via the associated extrinsic information geometry, the system looks as if it behaves as a functional of beliefs about external states.

## 4.3. Phenotypic representations? Ontologies?

Our last move is to leverage the fictionalist account to nuance the claim that the account just rehearsed vindicates the *neural representations* construct. This nuance is on two counts: with regards to the terms 'neural' and 'representations'.

First, the term 'neural' should be replaced by 'phenotypic', to reflect the expanded realization base of the vehicles of deflated, mathematical content under the free-energy principle. The spirit of deflated neural representationalism is vindicated by the free-energy principle. We can indeed assign mathematical content to structures that are internal to an agent, which come to encode or carry semantic content when the right ecological conditions are in play (thanks to the dual aspect information geometries of living systems). However, the internal states that play this role are very far from classical representations in the symbolic or connectionist traditions. They essentially comprise all internal states of a system and so are not strictly neural. This has the slight consequence that the neural representationalist intuition is vindicated by its traditional adversary, the embodied-enactive approach to cognition: if there is anything like structural representations under the free-energy principle, they correspond to the system's bodily states and are harnessed, maintained, and updated through histories of adaptive action.

Second, with respect to the term 'representation', we note that under the free-energy principle, the deflated representational structures may best be understood as the *ontology* that system brings to bear in understanding its environment; that is, the set of hypotheses or categories that it uses to parse the flow of its sensory states.

Heuristically, we say that the free-energy principle licenses the claim that the system believes that this or that environmental factor is causing its sensory impressions. In light of the discussion above, it appears more accurate to say that, when it is in its usual ecologically valid operating conditions, a system equipped with such a partition that exists at nonequilibrium steady state will act in a way that *looks as if* it has an intentional relation with some features of its environment. We now know what this "as if" character amounts to: it refers to the duality of information geometries and thereby the duality of possible descriptions (in terms of a flow towards nonequilibrium steady state and in terms of belief updating under a generative model).

The free-energy principle descends from a modeling strategy called generative modeling. In this scheme, we write down alternative probabilistic models of the process that caused our data, and we score the probability of each model as a function of how well it explains the variance in our data. This score is the variational free-energy. Crucially, as discussed in the previous section, these models are *hypotheses about the structure of the process that caused our data*. The hidden or latent (c.f., external) states in these models are essentially guesses about the causes of sensory data. Crucially, they need not reflect the existence of anything in reality (Baltieri and Buckley, 2019; Friston, 2020; Friston et al., 2020). This is a subtle but important point. The external states exist only relative to the generative model and accompanying phenotype, and only play a role so long as they subtend the generation of adaptive, contextually appropriate behavior.

What is being described by the formalism that underwrites the free-energy principle, then, is less a story about how an internal reconstruction of the external world is constructed in the brain, as traditional (symbolic and connectionist) accounts of neural representations would have it. What is at stake is more like the *ontology* with which a system is equipped (Kohn, 2013). An ontology, in this sense, is the set of semantic structures (a large part of which are learned through experience from immersion in specific contexts) that a living creature uses, implicitly or explicitly, to parse and order the flow of its sensory states (Kirmayer, 2018; Kirmayer and Ramstead, 2017; Ramstead et al., 2016; Veissière et al., 2020).

 

We are effectively combining the deflationary account and the fictionalist or instrumentalist account to provide an interpretation that might be called *organism-centered functionalism*. The organism's phenotype—its moving and acting body—is a nonequilibrium steady state density that can be interpreted as a manifold towards which the flow of system's states settles on average and over time; and also as a joint probability distribution over all the variables of the system. The organism's behavior is driven by these density dynamics: by the tendency to settle towards its nonequilibrium steady state density, which is implemented as a gradient descent on variational free-energy (a.k.a. active inference).

This is just another way of saying that the actions selected by the organism are driven by the statistical structure of its phenotype and interactions with the environment (Ramstead et al., 2019). The organism leverages its own statistical structure in driving its action selection. This dovetails with embodied-enactive approaches to cognition and effectively constitutes a new take on morphological, developmental, and evolutionary computation (Fields and Levin, 2020; Kuchling et al., 2019) and knowledge-driven skillful action (Hipolito et al., 2020).

While our aim here was to explore the consequences of non-realist approaches to cognition, we note that the pragmatist interpretation that we propose may be compatible with certain realist, structural representationalist accounts, where content is determined by functional isomorphism, thereby illuminating the role of structural representations using a form of functional role semantics (Kiefer and Hohwy, 2018, 2019; Kiefer, 2020). These accounts are explicitly developed in a direction that does not require the represented system to actually exist in reality, which coheres with the account of ontologies just presented. The free-energy principle warrants the claim that there are phenotypic states that carry semantic content; here, we articulated a computational theory proper for such mathematical semantics. What we call these states, at the end of the day, may be a matter of preference.

## 5. Conclusion

In this paper, we aimed to assess whether the construct of neural representations plays an explanatory role under the variational free-energy principle, and to determine which philosophical position about the ontological and epistemological status of the representations construct is most appropriate for that theory. We examined non-realist approaches, rather than the more commonly discussed realist ones. We started by a deflationary account of mental representation, according to which the explanatorily relevant contents of neural representations are mathematical, and a fictionalist account, according to which representations are scientifically useful fictions. We hope to have shown that under the free-energy principles, even quite minimalist, deflationary accounts of the kind of content carried by neural representations warrant an appeal to a semantic content, which echoes (while being distinct from) the realist position. We hope to have shown that, by drawing on a modified fictionalist account, the formal semantics derived from the free-energy principle can provide us with an explanation of the aboutness or intentionality of living systems.

Much hangs philosophically on what it means to represent some target domain, specifically in terms of the relation between mental states and the physical states that realize them. The relations between the free-energy principle and classical positions in the philosophy of mind (e.g., physicalist monism, dual-aspect monism, and Cartesian dualism) have been explored at length elsewhere (Friston et al., 2020). We will only comment on which of these seems to cohere most with our account. Briefly, of the philosophical perspectives on the relation between mental and physical states, ours here is most consonant with *functionalism* and the concept of *multiple realization* that it entails. Functionalism is the view that the features that characterized mental states are not the intrinsic features of that state, but rather the functional (e.g., input-output) relations between that state and other states of the system (Putnam, 1960, 1967). Multiple realization is the view that the same (mental) macrostate state can be realized variously by different configurations of (physical) microstates, so long as they implement the appropriate functional (e.g., computational) relations (Putnam, 1967). The proximity between our view and functionalism is based on the technical detail of how the semantic content is realized under the free-energy principle, via the assignment of fictive external states to internal states implied by the dual information geometry of the free-energy principle. As discussed above, the free-energy principle licences the claim that, for every blanket state, we can identify an *average of internal states* that we can associate with the parameters of a probability density over (fictive) external states. The crucial thing to note is that it is the *average* of internal states that can be so associated. This means that the probability density over external states can be realized by an equivalence class of internal states, which end of parameterizing the same belief.

An outstanding issue is whether the framework on offer is able to account for the issue of *misrepresentation*. Any candidate representational structure must at least in principle be able to misrepresent the state of affairs that it represents. This has long been a sticking point in the discussion on representation (Dretske, 1986). In a nutshell, because misrepresentation is possible (e.g., recognizing one object as another that it is not), an account of representation needs to allow for misrepresentation while also specifying what makes the representation about one object versus another – if it can be induced by both objects (Fodor, 1975, 2008). It has been pointed out, in our view correctly, that the information theoretic measures used under the free-energy principle are measures of mere covariance, which are insufficient to account for misrepresentation (Kirchhoff and Robertson, 2018). We think that this picture is incomplete. A future direction for more fully addressing this issue begins by noting that variational free-energy scores the degree to which sensory data conforms to hypotheses about what caused it. Variational free-energy is a not just any measure of information, but instead measures the discrepancy between the current sensory data and the sensory data *expected under some ontology or hypothesis*, which lends it an irreducibly semantic aspect. High free-energy indicates that the hypothesis does not 'explain' the data, or that some other hypothesis would fare better. Thus, the model of semantics on offer might be able to account for misrepresentation and the search for alternative hypotheses.

Our primary concern in this paper was to show that phenotypic states can come to acquire semantic contents of a deflationary (mathematical) sort via active inference. Of note is that this semantic content is not equivalent to the kind of *propositional* content that is at play in language use, nor does our account explain the manners in which human agents use language and narrative to fashion and remake themselves as agents (Bouizegarene et al., 2020; Taylor, 2016). While this issue is, at least arguably, beyond the scope of this paper, we generally agree with the claim that the move from semantic to propositional content requires that agents engage in specific kinds of content-involving practices. These are practices like truth-telling (Hutto and Myin, 2017; Hutto and Myin, 2013) and story-telling (Hutto, 2012), which build greatly upon the basic representational or semantic capacity of agents by enabling more elaborate, storied, forms of self-access that would not be possible without language.

Where does this leave us with regard to our initial questions? The upshot of our discussion is that, under the free-energy principle, there indeed are structures internal to an organism that act as the bearers of semantic content. These structures can be specified mathematically in terms of computational theory proper, as held by the deflationary account. However, by virtue of the dual information geometries in play under the free-energy principle, this purely mathematical account comes with an implicit semantics: the set of hypotheses about underlying causal factors (or the *ontology*) with which the system parses and makes sense of its sensory stream. This might be seen as vindicating the structural representationalist account discussed in the introduction, albeit with a critical twist: those structures that bear content are not merely *neural* representations, but indeed *phenotypic* representations (if they are representations at all), for it is all the internal states of an organism, given the Markovian partition, that bear this content.

**Technical appendix**

**1. The Langevin formalism and density dynamics**

One can express a random dynamical system in terms of the flow of states over time that are subject to random fluctuations (please see glossary for a definition of variables):

$$\dot{x}(\tau) = f(x, \tau) + \omega \tag{1.1}$$

This is a general specification of (Langevin) dynamics that underwrites nearly all of physics (Ao, 2008; Seifert, 2012; Sekimoto, 1998); in the sense that most modern physics is premised on the Langevin formalism and the ensuing descriptions of the flow of a system's states under random fluctuations. This can be equivalently described in terms of density dynamics – via the Fokker Planck equation or Schrödinger equation – or the path integral formulation. From these descriptions, nearly all quantum, statistical and classical mechanics can be derived.

We are interested in systems that have measurable characteristics, which means that they possess an attracting set or manifold, known as a random or pullback attractor (Crauel and Flandoli, 1994). This means that if we observe the system at a random time, there is a certain probability of finding it in a particular state. This is the nonequilibrium steady-state density (Seifert, 2012).

One can now use standard descriptions of density dynamics to express the flow of states as a gradient flow on *self-information* or *surprisal* (Jaynes, 1957; Jones, 1979; MacKay, 2003; Tribus, 1961). This flow is the steady-state solution to Fokker Planck equation that accompanies (1.1) (Frank, 2004; Frank et al., 2003; Kerr and Graham, 2000; Kim, 2018; Tomé, 2006).

$$f(x) = (Q - \Gamma) \cdot \nabla \Im(x)$$
$$\Im(x) = -\ln p(x) \tag{1.2}$$

This equation says that, on average, the states of any random dynamical system with an attracting set evince a gradient flow on surprisal; namely, the negative logarithm of the nonequilibrium steady-state density (Friston and Ao, 2012; Yuan et al., 2010). The gradient flow effectively counters dispersion due to random fluctuations, such that the probability density does not change over time. See Figure 4.

## The Langevin equation

$$\dot{x} = f(x) + \omega$$

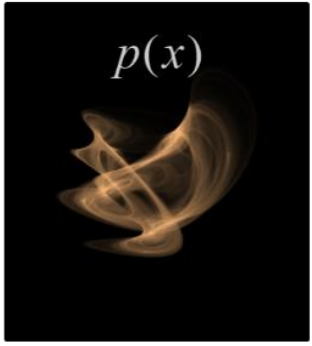

$p(x)$

## Nonequilibrium steady-state

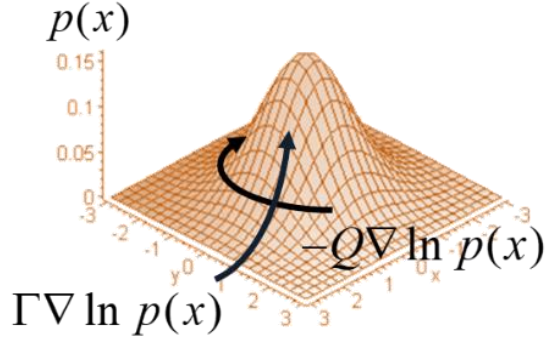

$p(x)$

$-Q \nabla \ln p(x)$

$\Gamma \nabla \ln p(x)$

The Fokker-Planck equation $\quad \dot{p}(x) = \nabla \cdot (\Gamma \nabla - f) p(x)$

And its solution $\quad \dot{p}(x) = 0 \Rightarrow f(x) = (Q - \Gamma) \nabla \Im(x)$

**Figure 6.** *Density dynamics and pullback attractors.* This figure depicts the density or ensemble dynamics of random dynamical systems that can be described via the Langevin equation. The left panel depicts the time evolution of two states, as a strange attractor. A point in this space assigns to the system a position along each dimension, and so assigns a value to each state. Here, each dimension represents one of the two states, and the trajectory plots the evolution of states over time. The right panel represents an arbitrary random attractor (a pullback attractor). One can think of this pullback attractor in two ways. First, the attractor can be cast as representing the trajectory of systemic states over time (in this case, two states are represented). The crucial feature of this trajectory is that – after sufficient time has passed – it will revisit specific regions of state space, which make up the pullback attractor itself. The second interpretation is probabilistic: it casts the attracting set as a probability density over the states in which the system can be found when it is sampled at random. The Fokker-Planck equation allows us to describe the evolution of this probability density. This, in turn, licences a solution to the Fokker-Planck equation. The consequence of this is that we can establish a lawful relationship between the probability density and the flow of states at any point in the system's state space. This solution describes the flow of systemic states in terms of gradients of log density or surprisal and in terms of the amplitude of random fluctuations. In turn, the Helmholtz decomposition allows us to express the nonequilibrium steady-state solution in terms of two orthogonal components. One of these is a curl-free gradient flow that depends on the amplitude of random fluctuations $\Gamma$. This component rebuilds probability gradients, effectively countering the effect of random fluctuations on states (i.e., countering their dispersion). The other component is a divergence-free (or solenoidal) flow that circulates on isoprobability contours and that depends upon an antisymmetric (skew) matrix $Q$. The figure depicts the flow around the peak of a probability density that has a Gaussian or normal form. See (Friston, 2019; Friston and Ao, 2012) for technical details.

The above equation holds (nontrivially) for the internal, blanket, and external states. If we just focus on internal and active (i.e., *autonomous*) states, we have the following flows. Note that as in (1.2) $Q_{\alpha\alpha}$ and $\Gamma_{\alpha\alpha}$ denote antisymmetric and leading diagonal matrices, respectively.

$$
\begin{aligned}
&f_\alpha(\pi) = (Q_{\alpha\alpha} - \Gamma_{\alpha\alpha}) \nabla_\alpha \Im(\pi) \\
&\alpha = \{a, \mu\} \\
&\pi = \{s, \alpha\} = \{b, \mu\}
\end{aligned}
\tag{1.3}
$$

This means anything with a Markov blanket must evince the above gradient flows. In turn, this means that internal and active states will look as if they are trying to minimise the same quantity; namely, the surprisal of states that constitute the thing, particle, or creature. These are the internal and blanket states, i.e., *particular states*.

## 2. Bayesian mechanics

If internal and external states are conditionally independent, then for every given blanket state there is an expected internal state and a conditional density over external states. In other words, there must be a one-to-one relationship between the average internal state of a particle (or creature) and a probability density over external states, for every given blanket state. This means that we can express the posterior or conditional density over external states as a *variational density* that is parameterised by internal states:

$$q_{\boldsymbol{\mu}}(\eta) = p(\eta \,|\, b) = p(\eta \,|\, \pi)$$
$$\boldsymbol{\mu}(b) \triangleq E[\mu \,|\, b] \tag{1.4}$$

This allows us to interpret the flow of autonomous states $\boldsymbol{\alpha} = \{a, \boldsymbol{\mu}\}$ (i.e., action and perception) as a gradient flow on variational free energy.

$$f_{\boldsymbol{\alpha}} = (Q_{\alpha\alpha} - \Gamma_{\alpha\alpha})\nabla_{\boldsymbol{\alpha}} F(b)$$

$$F(b) \triangleq \underbrace{E_q[\Im(\eta, b)]}_{\text{energy}} - \underbrace{H[q_{\boldsymbol{\mu}}(\eta)]}_{\text{entropy}} \tag{1.5}$$
$$= \underbrace{\Im(b)}_{\text{surprisal}} + \underbrace{D[q_{\boldsymbol{\mu}}(\eta) \,\|\, p(\eta \,|\, b)]}_{\text{bound}}$$
$$= \underbrace{E_q[\Im(b \,|\, \eta)]}_{\text{inaccuracy}} + \underbrace{D[q_{\boldsymbol{\mu}}(\eta) \,\|\, p(\eta)]}_{\text{complexity}} = \Im(b)$$

This functional can be expressed in several forms; namely, an expected energy minus the entropy of the variational density, which is equivalent to the self-information associated with blanket states (i.e., *surprisal*) plus the KL divergence between the variational and posterior density (i.e., *bound*), which, in this instance, is zero by (1.4). In turn, this can be decomposed into the expected log likelihood of blanket states (i.e., *accuracy*) and the KL divergence between posterior and prior densities (i.e., *complexity*).

The second thing that (1.4) brings to the table is an *information geometry* and attending calculus of beliefs. From now on, we will associate beliefs with the probability density above that is parameterised by (expected) internal states. Note that these beliefs are non-propositional, where 'belief' is used in the sense of 'belief propagation' and 'Bayesian belief updating' that can always be formulated as minimising variational free energy (Beal, 2003; Dauwels, 2007; Yedidia et al., 2005). To license a description of this conditional density in terms of beliefs, we can now appeal to information geometry (Amari, 1998; Ay, 2015; Caticha, 2015; Kim, 2018). Note the variational free energy—and its gradients—are functionals of a *generative model* $\Im(\eta, b) = -\ln p(\eta, b)$ in the form of a surprisal over external and blanket states. This means that the nonequilibrium steady-state density over the states can be read as a generative model that underwrites autonomous gradient flows.

### 3. Information geometry and beliefs

Any statistical manifold is necessarily equipped with a unique metric tensor, known as the Fisher information metric (Amari, 1998; Crooks, 2007; Kim, 2018).

$$
\begin{aligned}
d\ell^2 &= g_{ij} d\mu^i d\mu^j \\
g(\mu) &= \nabla_{\mu'\mu'} D[q_{\mu'}(\eta) \| q_\mu(\eta)]|_{\mu'=\mu} = E_q[\nabla_\mu \ln q_\mu(\eta) \times \nabla_\mu \ln q_\mu(\eta)]
\end{aligned}
\tag{1.6}
$$

Here, $d\ell$ is the information length associated with small displacements on the statistical manifold $d\mu = \mu' - \mu$ induced by a probability density $q_\mu(\eta)$. The information length scores the number of different probabilistic or *belief states* encountered in moving from one part of a statistical manifold to another.

If we return to the independencies induced by the Markov blanket, Equation (1.4) tells us something fundamental. The (expected) internal states have acquired an information geometry, because they parameterise probabilistic beliefs about external states. In short, there is a unique geometry in some belief space that can be associated with the internal (physical) state of any particle or creature. Furthermore, we know that the gradient flows describing the dynamics of internal states can be expressed as a gradient flow on a variational free energy functional (i.e., function of the function) *of beliefs*: see (1.5).

Recall from above, that an information geometry is a property of any statistical manifold. The parameters of the probability density over the internal states are thermodynamic variables $\lambda$ (e.g., pressure) that underwrite thermodynamics or statistical mechanics (Crooks, 2007; Holmes et al., 2019). We will refer to the accompanying information geometry as an *intrinsic* geometry, because pertains to the internal states per se. From our point of view, this means there are two information geometries in play with the following metrics:

$$
\begin{aligned}
g(\lambda) &= \nabla_{\lambda'\lambda'} D[p_{\lambda'}(\mu) \| p_\lambda(\mu)]|_{\lambda'=\lambda} \quad \text{intrinsic} \\
g(\mu) &= \nabla_{\mu'\mu'} D[q_{\mu'}(\eta) \| q_\mu(\eta)]|_{\mu'=\mu} \quad \text{extrinsic}
\end{aligned}
\tag{1.7}
$$

First, there is an *intrinsic* information geometry based upon thermodynamic variables. This forms the basis of statistical mechanics in physics. At the same time, there is an information geometry in the space of internal states that refers to belief distributions over external states. This is the *extrinsic* information geometry that inherits its properties from the *Markovian* conditions that define, stipulatively, autonomous states (via their Markov blanket). The extrinsic geometry is conjugate to the intrinsic geometry, in the sense that they supervene on the same Langevin dynamics.

## Glossary of terms and expressions

(Note: a.u. stands for arbitrary units, e.g., metres (m), radians (rad), etc.)

| Expression | Description | Units |
|---|---|---|
| **Variables** | | |
| $\omega(\tau)$ | Random fluctuations | a.u. (m) |
| $x = \{\eta, s, a, \mu\} \in X$ | Markovian partition into external, sensory, active, and internal states | a.u. (m) |
| $\alpha = \{a, \mu\} \in A$ | Autonomous states | a.u. (m) |
| $b = \{s, a\} \in B$ | Blanket states | a.u. (m) |
| $\pi = \{b, \mu\} \in P$ | Particular states | a.u. (m) |
| $\eta \in E$ | External states | a.u. (m) |
| $\Gamma = \mu_m k_B T$ | Amplitude (i.e., half the variance) of random fluctuations | J·s/kg |
| $Q$ | Rate of solenoidal flow | J·s/kg |
| $\ell = \int d\ell : d\ell^2 = g_{ij} d\lambda^j d\lambda^i$ | Information length | nats |
| $g_{ij} = E\left[\dfrac{\partial \mathfrak{I}}{\partial \lambda^i} \dfrac{\partial \mathfrak{I}}{\partial \lambda^j}\right]$ | Fisher (information metric) tensor | a.u. |
| **Functions, functionals and potentials** | | |
| $E[x] = E_p[x] = \int x p_\lambda(x) dx$ | Expectation or average | |
| $p_\lambda(x) : \Pr[X \in A] = \int_A p_\lambda(x) dx$ | Probability density function parameterised by sufficient statistics $\lambda$ | |
| $q_\mu(\eta)$ | Variational density – an (approximate posterior) density over external states that is parameterised by internal states | |
| | | |
| $F(b) \geq \mathfrak{I}(b)$ | Variational free energy free energy – an upper bound on the surprisal of particular states | nats |
| **Operators** | | |
| $\nabla_x \mathfrak{I}(x) = \dfrac{\partial \mathfrak{I}}{\partial x} = \left(\dfrac{\partial \mathfrak{I}}{\partial x_1}, \dfrac{\partial \mathfrak{I}}{\partial x_2}, \cdots\right)$ | Differential or gradient operator (on a scalar field) | |
| **Entropies and potentials** | | |
| $\mathfrak{I}(x) = -\ln p(x)$ | Surprisal or self-information | nats |
| $D[q(x) \parallel p(x)] = E_q[\ln q(x) - \ln p(x)]$ | Relative entropy or Kullback-Leibler divergence | nats |

## References (Technical appendix and Glossary of terms and expressions)